# Yttrium Iron Garnet Thickness Influence on the Spin Pumping in the Bulk Acoustic Wave Resonator

S.G. Alekseev[a,*], N.I.Polzikova[a], and A.O.Raevskiy[b]

[a] *Kotel'nikov Institute of Radioengineering and Electronics of Russian Academy of Sciences,*
*125009 Mokhovaya str. 11, build. 7, Moscow, Russia*
[b] *Fryazino branch Kotel'nikov Institute of Radioengineering and Electronics of Russian Academy of Sciences,*
*141190 Vvedenskiy sq.,1, Fryazino, Moscow region,  Russia*
*E-mail:* * *alekseev@cplire.ru*

*Abstract* -- The features of phonon – magnon interconversion in acoustic resonator determine the efficiency of spin pumping from YIG into Pt that may be detected electrically through the inverse spin Hall effect (ISHE). Based on the methods developed in previous works for calculating resonator structures with a piezoelectric (ZnO) and a magnetoelastic layer in contact with the heavy metal (YIG/Pt), we present the results of numerical calculations of YIG film thickness influence on acoustically driven spin waves. We obtain some YIG film thickness regions with various behavior of dc ISHE voltage $U_{ISHE}$. At micron and submicron thicknesses, the higher spin wave resonance (SWR) modes (both even and odd) can be generated with efficiency comparable and even exceeding that of the main mode. The absolute maximum of $U_{ISHE}$ is achieved   at the thickness about $s_1 \approx 208$ nm under the excitation of the first SWR.

## Introduction

With the development of new materials and structures, as well as the development of experimental techniques, various aspects of the magnon-phonon interaction acquired a new sense and are currently being actively explored in microwave spintronics, streintronics, spincaloritronics and magnonics [1-8]. The piezoelectric generation of acoustically driven spin waves (ADSW) in piezoelectric /magnetoelastic structures is promising for use in low power consumption devices free from ohmic losses [7-12].  In particular, acoustic spin pumping (ASP) – the generation of spin polarized electron currents from ADSW [6,7] – is promising for application in microwave spintronics and attracts much attention of the researchers. In addition, the electrical detection of magnons by means of the ASP and the inverse spin Hall effect (ISHE) is a useful method for investigations of magnetization dynamics and phonon – magnon interconversion.

The magnon – phonon interactions in ferrimagnetic yttrium iron garnet (YIG) films on gallium gadolinium garnet (GGG) substrates have been studied for many decades. The main attention was paid to the study of lateral wave propagation both under the excitation of magnetic precession and under the acoustic wave (AW) excitation   by different piezoelectric transducers [5, 13-16]. The magnetoelastic interaction in thickness mode resonators of bulk AW with ferromagnetic and piezoelectric films was investigated in our previous works [10-12].

In [11] we experimentally demonstrated the excitation of ADSW by means of high overtone bulk AW resonator (HBAR) containing YIG and piezoelectric zinc oxide (ZnO) films.  The resonant ASP in HBAR containing YIG/Pt system was proposed and implemented in [17,18]. In [19,20], a good agreement between the theoretical and experimental frequency-field dependences $U_{ISHE}(f, H)$ of the magnitude of the ISHE dc voltage was shown.

The effect of geometry  on ISHE induced by the spin pumping driven by the ferromagnetic resonance (FMR) or by spin waves (SW)  have been investigated in many studies (see, for example, [21- 23]). The effect of magnetoelastic film thickness on power absorption in FMR driven  by surface AW was studied in [8].

Here, we report on the theoretical investigation of the influence of the YIG thickness on the phonon – magnon interconversion efficiency in an acoustical cavity and hence, on the ASP efficiency. In particular, we study the features of the excitation (by means of HBAR) and detection (through the ASP and ISHE) of SW resonances (SWR) in micron and nanometer-sized YIG films.

## 1. HBAR geometry

The structure under consideration and main vectors orientation are shown in Fig. 1. A transducer consisting of a piezoelectric ZnO film *1*, sandwiched between thin-film Al electrodes *2*, is deposited on the top of the GGG substrate - YIG film structure *3-4*. A thin Pt strip *5* is attached to the YIG film underneath the acoustic resonator aperture. The external magnetic field $\vec{H}$ lies in the plane of the structure along the *z*-axis and magnetizes YIG films up to uniform saturation magnetization $\vec{M}_0 \parallel \vec{H}$. To excite the bulk AW propagating along the *x*-axis perpendicular to the layers, the rf voltage $\tilde{U}(f)$ with frequency *f* is applied across the transducer.

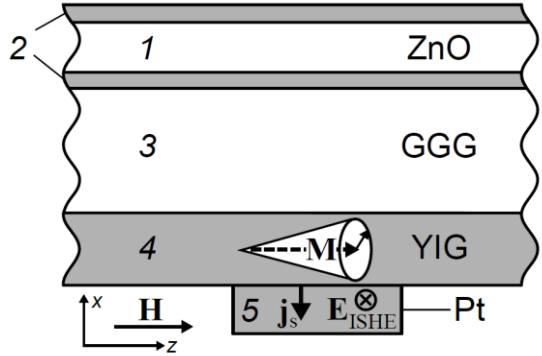

**Fig. 1.** The layout of a bulk acoustic wave resonator.

It is assumed that ZnO film excites shear bulk AW with polarization along the magnetic field $\vec{H}$ [24, 25]. In the YIG layer, this wave drives magnetization dynamics due to the magnetoelastic interaction. These ADSWs establish a spin current $\vec{J}_s$ from YIG into the Pt strip. The ISHE converts $\vec{J}_s$ to a conductivity current or an electrostatic dc field $\vec{E}_{ISHE}$.

The typical thickness parameters for the GGG substrate (*d*=500μm), piezoelectric ZnO (*l*=3μm), and platinum 12 nm are used for the calculations. The positions of the interfaces YIG/Pt and YIG/GGG are $x = x_0$ and $x = x_0 + s$, respectively. The YIG thickness *s* is varied from some nanometers to several ten microns.

## 2. Method of Calculation

Let us consider the features of ISHE induced by ASP in our resonator structure. The dc voltage between Pt ends is

$$U_{ISHE} = -(\vec{E}_{ISHE} \cdot \vec{a}) \propto \theta^2 ((\vec{n} \times \vec{z}) \cdot \vec{a}).  \quad (1)$$

Here,

$$\theta = \sqrt{\text{Im}[m_x^*(x_0) m_y(x_0) / M_0^2]},  \quad (2)$$

is the magnetization $\vec{M} = \vec{M}_0 + \vec{m}$ precession cone angle at YIG/Pt interface $x = x_0$, $\vec{m} = (m_x, m_y, 0)$ is the ac magnetization, $\vec{n}$ is the normal to the interface, and $\vec{a}$ is the vector in the direction of the Pt strip [26, 27]. In (1) we omit the characteristics of the spin detector (Pt) and YIG/Pt interface: the spin Hall angle and the spin mixing conductance, because they are considered to be independent of the magnetic field, frequency, and thickness of YIG and Pt. We also omitted the factor resulting from the current density averaging across the Pt thickness. So we calculate the ISHE voltage dependency on the YIG thickness while keeping all other parameters constant.

To find the components of the ac magnetization $m_x$ and $m_y$, entering into Eq. (2), we use one-dimensional and linear approximations, assuming that all variables depend on coordinate $x$ and time $t$ as $\exp[i(k^{(j)}x-2\pi ft)]$, where $k^{(j)}$ is the wavenumber in the $j$-th layer. In all nonmagnetic layers, $k^{(j)} = \pm(2\pi f/V^{(j)})$, where $V^{(j)}$ is the phase velocity of a shear AW.

In ferrite, magnetoelastic waves obey the secular equation [12, 28, 29]:

$$(f^2 - f_{AW}^2)(f^2 - f_{SW}^2) = \xi f_H f_M f_{AW}^2. \qquad (3)$$

Here, the terms in the brackets represent the dispersion laws for noninteracting AW and SW;

$$2\pi f_{AW} = kV, \quad f_{SW}^2 = f_H(f_M + f_H),$$

$$f_M = \gamma 4\pi M_0, \quad f_H = \gamma(H + H_{ex}), \quad H_{ex} = Dk^2, \qquad (4)$$

$D$, $H_{ex}$ are the exchange stiffness and magnetic field, $\gamma \approx 2.8$ MHz/Oe is the gyromagnetic ratio, and $\xi = b^2/(4\pi CM_0^2)$ is the parameter of AW and SW interaction, depending on effective magnetoelastic constant $b$ [30, 31] and the elastic modulus $C$. Note, that the inclusion of the exchange field $H_{ex}$ to (3), (4) leads to the increase of the equation order in $k$ (from biquadratic to bicubic). This significantly complicates the whole problem, in particular demands to use additional boundary conditions for magnetization. Nevertheless, we take $H_{ex}$ into account because of its significance for the formation of coupled magnetoelastic wave spectrum and efficiency of magnon – phonon interconversion [9, 12, 28]. The additional boundary conditions are taken here in the form (so called free surface spins)

$$\left.\frac{\partial m_{x,y}(x)}{\partial x}\right|_{\substack{x=x_0 \\ x=x_0+s}} = 0. \qquad (5)$$

For further calculations, we use the self-consistent method accounting for the back action of ADSW in YIG films on the elastic subsystem in all layers of the structure (in non-magnetic layers through boundary conditions) [19, 20]. The following elastic and magnetic parameters are used: (111) oriented YIG – $V^{(4)} = 3.9\times10^5$ cm/s, $\rho^{(4)} = 5.17$ g/cm$^3$, $b = 4 \times 10^6$ erg/cm$^3$, $D = 4.46\times10^{-9}$ Oe cm$^2$, GGG – $V^{(3)} = 3.57\times10^5$ cm/s, $\rho^{(3)} = 7.08$ g/cm$^3$ [13, 29]; ZnO – $V^{(1)} = 2.88\times10^5$ cm/s, $\rho^{(1)} = 5.68$ g/cm$^3$, where $\rho^{(j)}$ is the mass density of $j$-th layer. We take $4\pi M_0 = 955$ G as the characteristic value for La, Ga-substituted YIG epitaxial films used in the experiment [18]. The magnetic losses are taken into account by replacing in Eqs. (3) and (4) $f_H \rightarrow f_H + i\gamma\Delta H$, where $\Delta H = 0.70$ Oe is the FMR linewidth. For taking into account the elastic losses, the analogous replacing for elastic moduli $C^{(j)} \rightarrow C^{(j)} + i2\pi f\eta^{(j)}$, where $\eta^{(j)}$ is the viscosity factor, is performed for all media.

## 3. Results and discussions

The normalized frequency depenedencies $U_{ISHE}(f)$ for different YIG film thicknesses are represented in Fig.2. Here all data are normalized to the maximal voltage for the film of 31 μm thickness (see Fig.2a) used in the experiment [18]. All the dependencies are calculated for magnetic field value 740 Oe. The dashed-dotted lines point to the FMR frequency $f_{FMR}=f_{SW}(k=0) = 3.113$ GHz. The dashed lines correspond to the magnetoelastic resonance (MER) frequency $f_{MER}$, determined by the crossover of the dispersion dependencies of noninteracting AW and SW. The difference $f_{MER} - f_{FMR} \approx$ ≈ 30 MHz is due to the exchange interaction. Being two orders of magnitude lower than the FMR frequency, this difference turns out to be significant, since it exceeds by an order of magnitude the frequency difference of two adjacent HBAR resonances $f_{m+1} - f_m \sim 3$ MHz.

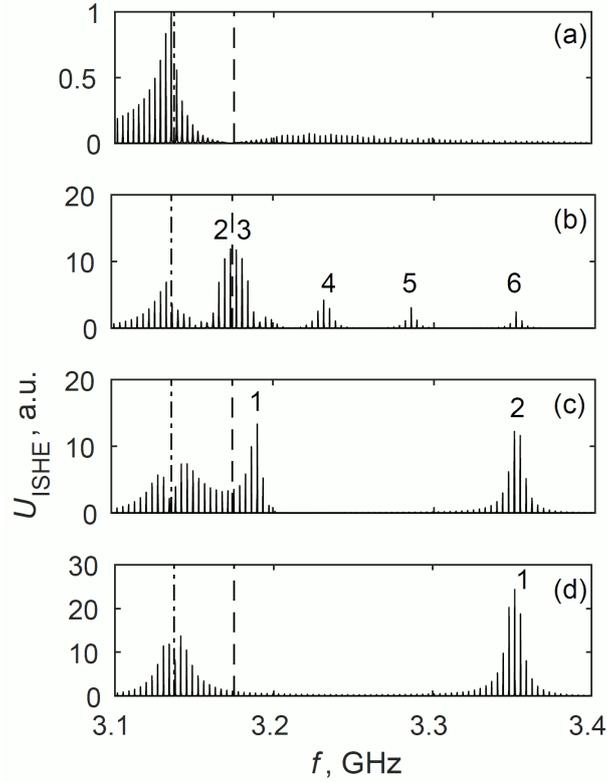

**Fig. 2.** Frequency dependencies at fixed magnetic field $H_0 = 740$ Oe of normalized dc voltages $U_{ISHE}$ on Pt stripe for structures with YIG thicknesses: $s = 31$ $\mu$m (a), $s = 1.5$ $\mu$m (b), $s = 0.5$ $\mu$m (c), and $s = 0.25$ $\mu$m (d). Dashed line - frequency of magnetoelastic resonance ($f_{MER}(H_0)$), dashed-dotted -- frequency of ferromagnetic ($f_{FMR}(H_0)$) resonances.

The ISHE voltage frequency dependence has the form of narrow resonances corresponding to AW resonances $f_m$ in the whole HBAR structure. The examples of such correspondence are shown in [19,20]. For every thickness $s$, one of the local maxima of the $U_{ISHE}(f)$ envelope is localized at the frequency $f_0 \approx f_{FMR}$ as one can see from Fig.2. Let's call $U_{ISHE}(f_0, s)$ the main maximum at the main frequency.

In Fig. 3, the dashed-dotted line represents the dependence of $U_{ISHE}(f_0, s)$ versus YIG thickness. With the increase of $s$ from several nanometers the value of $U_{ISHE}(f_0, s)$ increases approximately as $s^2$ and reaches a maximum for the thickness about $s_0 \sim 140$ nm. With the further increase of $s$, $U_{ISHE}(f_0,s)$ decreases, oscillating with the period $\sim 0.65$ $\mu$m, which corresponds to the AW half-length in YIG at frequency $f_0$.

As can be seen from Fig.2b-2d apart the main frequency there are additional frequency ranges of ADSW excitation localized at frequencies $f_n > f_0$ ($n = 1, 2, 3, ...$), dependent on the YIG thickness $s$. It can be seen that the frequencies of the SW resonances (SWR) $f_{SW}(k = k_n = \pi n/s)$ (marked with dots on the insertion in Fig.3), and the frequencies $f_n$ approximately coincide each other

$$f_n(s) \approx f_{SW}(k = k_n = \pi n/s). \qquad (6)$$

Note that all SW modes with the frequencies $f_n > f_{MER}$ have wave numbers exceeding the wave number of AW at that frequency. For example, the SWRs with $n>3$ in the inset have $k > 0.5 \cdot 10^7$ m$^{-1}$.

The thickness dependences of maxima of $U_{ISHE}(f_n,s)$ created by SWRs are shown in Fig. 3 by the curves 1-6. The curve numbers correspond to the mode numbers in Fig.2. As one can see, the $U_{ISHE}(f_n,s)$ values may exceed the voltage at the main frequency $U_{ISHE}(f_0,s)$. The maximum voltage of each mode is presented in Fig.3 only for those $s$ when the mode is clearly identifiable and not merged with other modes. For example, in Fig. 2b the mode $n=1$ is merged with the main mode, the modes $n=2$ and $n=3$ are also merged together. With the decrease of $s$, the modes with lower $n$ became clearly resolved (see Fig. 2c,2d) and the corresponding $U_{ISHE}(f_n,s)$ increases. At the thickness about $s_1 \approx 208$ nm, for the SWR mode with $n=1$, the value $U_{ISHE}$ reaches its absolute maximum.

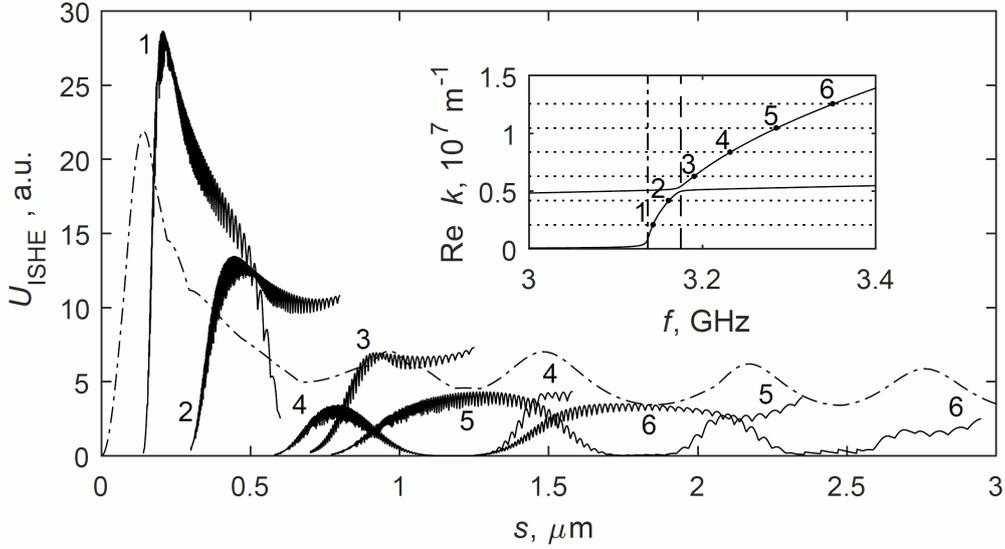

**Fig. 3.** Thickness dependencies at fixed magnetic field $H_0 = 740$ Oe of the main maximum $U_{ISHE}(f_0,s)$ (dashed-doted curve) and $U_{ISHE}(f_n,s)$ (curves 1-6). On the insertion: solid curves - obtained from Eqs. (3) and (4) frequency dependencies Re $k(f)$; horizontal dotted lines - SWR wave numbers $k_n = \pi n/s$ for $s=1.5$ μm and $n=1$-6; vertical dashed and dashed-dotted lines - $f_{MER}(H_0)$ and $f_{FMR}(H_0)$ as in Fig.2.

It is interesting to note that in contrast to the excitation of SWRs with microwave uniform magnetic field in our case the effective magnetic field is created by the elastic strain of AW and significantly non-uniform in the film thickness. As a result, the conditions are created for the excitation of both even and odd SWR modes. Nevertheless, at some $s$ values, the excitation of either even or odd modes takes place.

So SWR modes manifest themselves only in a certain range of thicknesses of YIG films, which in turn depends on the material parameters of YIG. The thickness decrease leads to the increase of frequency $f_n$ and its distance from MER. As one can see from Fig.3 the lower thickness value for ASDW resonances with $n \geq 1$ is about 140 nm.

Note, that in our calculations we intentionally do not consider the dependency of material parameters, for example, the magnetic damping, on YIG thickness. So, the thickness influence on $U_{ISHE}$ is entirely determined by the method of the SW excitation. Taking the additional magnetic damping due to the spin pumping which depends on thickness [26] into account leads to a steeper curve $U_{ISHE}(f_0,s)$ for the small $s < 200$ nm [22].

## Conlusion

The carried out studies show the strong influence of magnetoelastic film thickness on acoustic spin pumping in HBAR with ZnO-GGG-YIG/Pt structure.

Due to the inhomogeneous character of the exciting effective magnetic field of an elastic origin, higher SWR modes (both even and odd) can be generated with an efficiency comparable and even exceeding that of the main mode. The absolute maximum of $U_{ISHE}(f_1, s_1)$ is located at the frequency of the first SWR mode and at the thickness $s_1 \approx 208$ nm. Supposing that the magnitude of the normalization voltage $U_{ISHE}(f_0, 31\ \mu m) \sim 4\ \mu V$, which was observed in the experiment [18] it is expected that for the films with the optimal thickness $s \sim s_1$ and for the same applied power the $U_{ISHE}(f_1, s_1)$ value will be about hundreds of microvolts.

## Acknowledgments

This work was carried out within the framework of the state task and partially was supported by grant 17-07-01498 from the Russian Foundation for Basic Research.